\documentclass[11pt,aps,prl,reprint,superscriptaddress,floatfix]{revtex4-2} 
\usepackage{braket} 
\usepackage{url} 
\usepackage{cancel}
\usepackage{siunitx}    

\usepackage{color}
\usepackage{multirow}
\usepackage{bm}
\usepackage{epsfig}
\usepackage{psfrag}
\usepackage[latin1]{inputenc}
\usepackage[english]{babel}
\usepackage{amsfonts}
\usepackage{amsmath}
\usepackage{upgreek}
\usepackage[hidelinks]{hyperref}
\usepackage[dvipsnames]{xcolor}
\usepackage[normalem]{ulem}
\usepackage{bbold} 

\newcommand{\GammaF}{\varGamma} 

\newcommand{\td}{{\rm d}}
\newcommand{\e}{{\rm e}}
\newcommand{\ex}[1]{\langle #1 \rangle}	

\newcommand{\disAD}{{b}}

\usepackage{pdfpages} 
\usepackage{pgffor} 

\makeatletter
\AtBeginDocument{\let\LS@rot\@undefined}
\makeatother

\def\supplementfilename{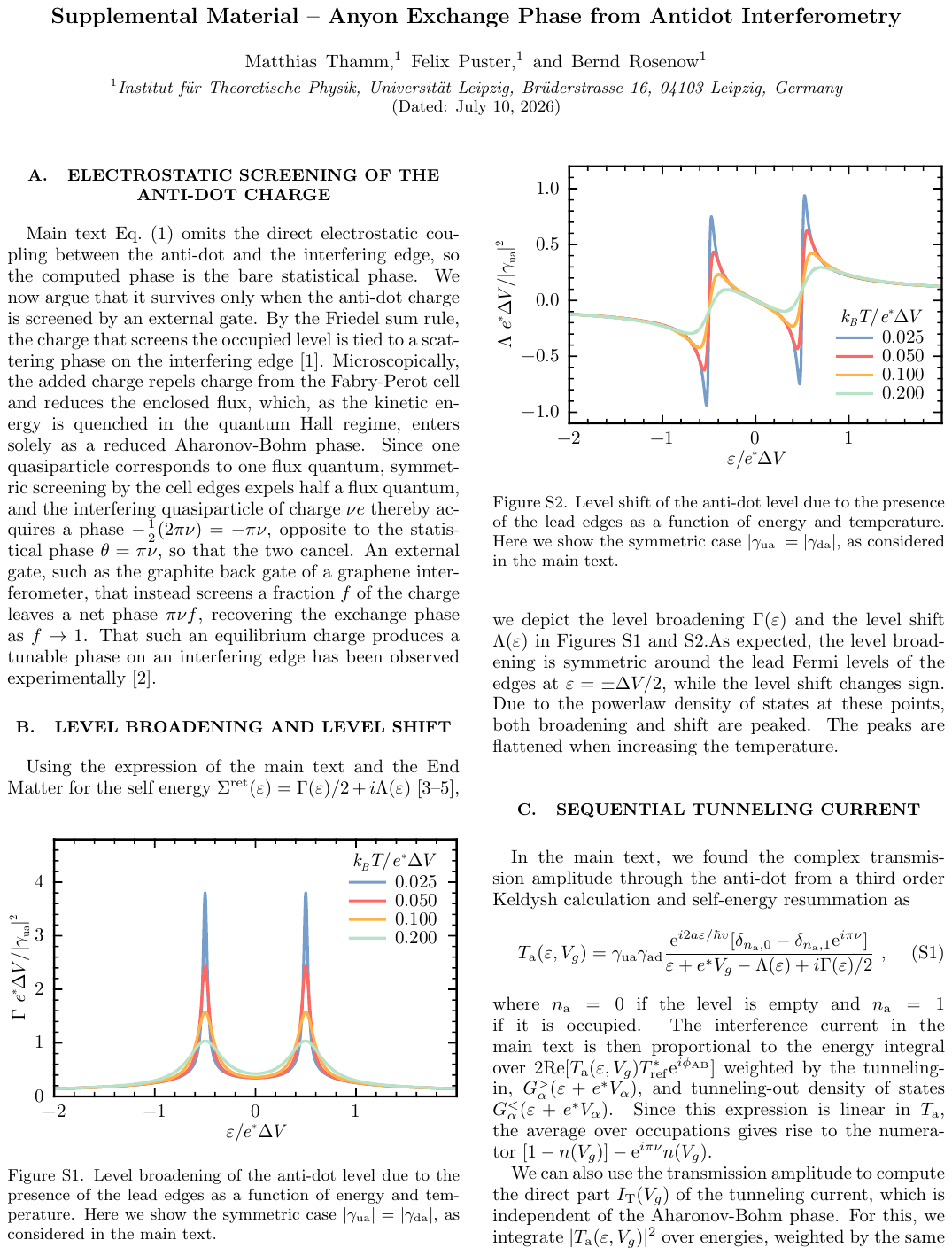}

\pdfximage{\supplementfilename}
\def\numbersupplementpages{\the\pdflastximagepages}

\newif\ifarXiv
\arXivtrue 

\begin{document}
	\title{Anyon Exchange Phase from Antidot Interferometry}
    
    \author{Matthias Thamm} 
	\email{thamm@itp.uni-leipzig.de}
	\affiliation{Institut f\"ur Theoretische Physik, Universit\"at Leipzig, Br\"uderstrasse 16, 04103 Leipzig, Germany}
    
	\author{Felix Puster}
	\affiliation{Institut f\"ur Theoretische Physik, Universit\"at Leipzig, Br\"uderstrasse 16, 04103 Leipzig, Germany}

	\author{Bernd Rosenow}
	\affiliation{Institut f\"ur Theoretische Physik, Universit\"at Leipzig, Br\"uderstrasse 16, 04103 Leipzig, Germany}

    \date{\today}

    \begin{abstract} Quasiparticles in fractional quantum Hall systems are anyons, carrying a fraction of the electron charge. Exchanging two of them gives rise to a fractional exchange phase. While the fractional charge and the braiding phase---twice the exchange phase---have been measured, the exchange phase itself has remained inaccessible. We study a quantum antidot embedded in a Fabry-Perot interferometer. Within a systematic non-equilibrium Keldysh treatment that consistently includes the occupation and level broadening of the antidot, we find that the transmission phase evolves non-monotonically when a gate voltage tunes the antidot through a resonance, in contrast to the monotonic evolution for electrons. The bare exchange phase can be extracted from the difference between the phase plateaus. 
    \end{abstract}
    \maketitle

\emph{Introduction---}
In three spatial dimensions only bosons and fermions exist, with exchange phases $0$ and $\pi$.  Two dimensions additionally allow anyons with intermediate exchange phase $\theta$~\cite{Leinaas.1977,Laughlin.1983,Halperin.1984,Arovas.1984,Feldman.2021}. Fractional quantum Hall systems at filling fraction $\nu$ host such anyons, characterized by a fractional charge $e^*$ and an exchange phase $\theta$~\cite{Halperin.1984,Arovas.1984}. The fractional charge has been measured in shot-noise experiments~\cite{Kane.1994,Reznikov.1999,Saminadayar.1997,dePicciotto.1997}, and the braiding phase $2\theta$---acquired when one anyon encircles another---is accessible in interferometer~\cite{C.Chamon.1997,Halperin.2011,Rosenow.2020,Mross.2026} and anyon-  collider~\cite{Rosenow.2016,Thamm.2024,Iyer.2024,Schiller.2023,Ronetti.2025a,Ronetti.2025b,Demazure.2026} schemes based on time-domain braiding~\cite{Han.2016,Lee.2019,Lee.2020,Schiller.2022,Rosenow.2025}.  The fractional braiding phase has been observed by several groups~\cite{Nakamura.2020,Nakamura.2022,Nakamura.2023,Bartolomei.2020,Lee.2023,Ruelle.2023,Glidic.2023,Kundu.2023,Ghosh.2025a,Ghosh.2025b,Samuelson.28.03.2024,Kim.27.12.2024,Werkmeister.2025}. The exchange phase $\theta$ itself, however, is thereby determined only modulo $\pi$~\cite{Read.2024}.

Proposals to access $\theta$ directly either fit interference data to theoretical predictions~\cite{Safi.2001,Kim.2005,Kim.2006,Vishveshwara.2003,Campagnano.2012,Campagnano.2013,Varada.2025}---where $\theta$ appears alongside non-universal parameters, is not directly measurable, or requires experimentally challenging regimes---or detect phase shifts in interference measurements~\cite{Kivelson.2024a,Puster.2025}, a more direct route. Kivelson and Murthy~\cite{Kivelson.2024a} proposed an interferometer in which tunneling at one junction proceeds through a near-resonant antidot level: tuning the level through resonance switches transport from direct tunneling through the empty level to co-tunneling through the occupied one, the two processes differing by $\theta$. They establish this for bosons and fermions in a single-level model; for anyons, their semiclassical and Berry-phase analyses cannot, as the authors note, fully separate the statistical phase from geometric contributions such as a change of the effective interferometer area, and a microscopic transport theory has been lacking.

\begin{figure}
    \centering
    \includegraphics[width=1\linewidth]{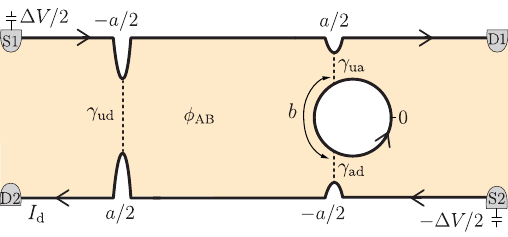}
    \caption{\label{fig:1setup}
    Fabry-Perot interferometer with a quantum antidot (a) of length $L$ embedded in one arm. The upper (u) and lower (d) edges, held at biases $+\Delta V/2$ and $-\Delta V/2$, are connected by a quantum point contact (QPC) with tunneling  amplitude $\gamma_{\rm ud}$, and the antidot couples to both edges through two further QPCs with tunneling amplitudes $\gamma_{\rm ua}$ and $\gamma_{\rm ad}$, separated by a distance $\disAD$ on the antidot and by $a$ along each edge. The current $I_{\rm d}$ is measured in drain D2.  }
    \end{figure}
In this Letter, we provide such a theory for the setup of
Ref.~\cite{Kivelson.2024a}, a Fabry-Perot interferometer with a quantum antidot embedded in one arm and tuned through a resonance by a plunger gate voltage $V_g$ at finite bias $\Delta V$ (Fig.~\ref{fig:1setup}). An electronic $T$-matrix benchmark clarifies how the exchange phase enters: the $0\to\pi$ evolution through the resonance is the familiar Breit-Wigner phase of an empty level with no co-tunneling, and for fermions the exchange factor $e^{i\pi}=-1$ is cancelled by the sign of the virtual intermediate-state energy. For Laughlin anyons~\cite{Laughlin.1983} the fractional factor $e^{i\pi\nu}$ survives as the measurable plateau difference.  

Our perturbative Keldysh calculation carries the anyonic statistics in the Klein factors and identifies the plateau difference with the bare exchange phase $\theta = \pi \nu$.  At fixed interferometer area this difference is purely statistical, and an electrostatic analysis identifies the gate-screened regime in which the area stays fixed \cite{Supplement}, isolating the exchange phase from the change in enclosed area, the dominant geometric effect noted in Ref.~\cite{Kivelson.2024a}. Treating the fractional edges as chiral Luttinger liquids, we include the power-law tunneling density of states, the resulting broadening of the anti-dot level, and its occupation out of equilibrium, and obtain the full gate-voltage dependence of the transmission phase. It evolves non-monotonically near the lead chemical potentials $V_g=\pm\Delta V/2$ (Fig. 2), in contrast to the monotonic $0 \to \pi$  evolution for electrons.

While the naive readout of $\theta$ thus fails near $V_g=\pm\Delta V/2$, the exchange phase $\theta=\pi\nu$  remains encoded in the difference between the transmission-phase plateaus on either side of a single resonance, robust against finite temperature and small geometry changes provided the antidot charge is gate-screened. Tuning between consecutive levels, by contrast, is dominated by geometry-dependent dynamical phases and is unsuitable for extracting $\theta$. The non-monotonic evolution itself is a direct prediction
for the ongoing experiment~\cite{Ehrets.2025}.

\emph{Setup---}We consider the Fabry-Perot interferometer of
Fig.~\ref{fig:1setup}, with a quantum antidot in one arm coupled to the upper (u) and lower (d) edges, which are themselves connected by a quantum point contact (QPC). Anyons tunnel at the QPCs according to
\begin{align}
        H_{\rm tun} 
        = &-\gamma_{\rm ua}\psi_{\rm a}^\dagger(x_{\rm u})\psi_{\rm u}\!\left(\frac{a}{2}\right)-\gamma_{\rm ad}\psi_{\rm d}^\dagger\!\left(-\frac{a}{2}\right)\psi_{\rm a}(x_{\rm d}) 
        \notag \\ & 
        -\gamma_{\rm ud} \psi_{\rm d}^\dagger\!\left(\frac{a}{2}\right) \psi_{\rm u}\!\left(-\frac{a}{2}\right)  + \rm h.c. \ , \label{Eq:Htun1}
    \end{align}
where $\gamma_{ij}=\gamma_{ji}^*$ is the complex tunneling amplitude from edge $i$ to $j$, $a$ the distance between the QPCs on the u and d edges, and $\psi_i^\dagger(x)$ creates an anyon
at position $x$ in channel $i$. The tunneling points on the antidot are a distance $\disAD=x_{\rm u}-x_{\rm d}$ apart. The antidot is a finite periodic edge of length $L$,
\begin{align}  
    H_{0,\rm a} &= \hbar v_{\rm a}\sum_{q>0} q b_q^\dagger b_q + \hbar v_{\rm a}\frac{\nu\pi}{L}  N(N-1)-e^*V_{g}N \ , \label{Eq:HamiltonianH0a}
\end{align}
with plunger gate $V_g$ setting the level positions and occupation, plasmon operators $b_q$ at momenta $q=2\pi n/L$ and velocity $v_{\rm a}$, and number operator $N$. The equilibrium chemical potential is at zero and the second term gives the Coulomb blockade. The edges u and d have velocity $v\neq v_{\rm a}$.

\begin{figure}
        \centering
        \includegraphics[width=1\linewidth]{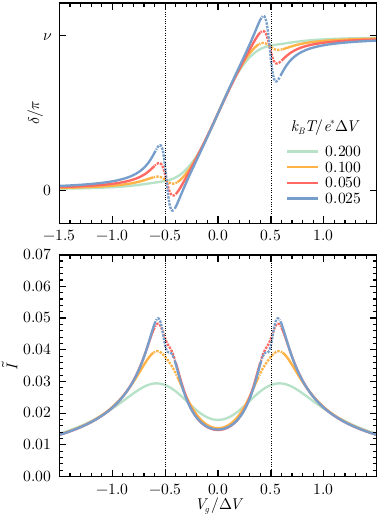}
        \caption{\label{fig:3temperaturedependencelargeT}
        Transmission phase $\delta$ (upper panel) and oscillation amplitude $\tilde{I}$ (lower panel) for tunneling through a single antidot level, versus gate voltage $V_g$ at several temperatures, for $\nu=1/3$ anyons and $a=0$. The Aharonov-Bohm part of the current is $I_{\rm d, AB} \propto \tilde{I}\cos(\phi_{\rm AB}+\delta+\pi)$ [Eq.~\eqref{Eq:FullPhaseTNonEq}], where the phase shift of $\pi$ amounts to the Breit-Wigner phase below resonance, and $\Delta V$ is the bias between edges u and d. {Regions in which the squared renormalized coupling at infrared scale, $r$, exceeds $0.5$ are shown with dashed lines. In these regions, the lineshape is qualitative rather than quantitative.} The interference current is largest near the edge chemical potentials $V_g=\pm\Delta V/2$ (dotted vertical lines), where the phase evolves non-monotonically.
}
    \end{figure}
    
\emph{Electron $T$-matrix calculation---}We first consider electrons, $\nu=1$, for which the transmission phase through a resonance increases monotonically from $0$ to $\pi$---the familiar Breit-Wigner phase of an empty-band approximation \cite{schuster1997phase}. Since this approximation involves neither an occupied level nor co-tunneling \cite{konig2001coherence,konig2002aharonov}, it is not obvious how an exchange phase can enter, which we now clarify.

Energies are measured from the common chemical potential of the edges and the antidot. For large level spacing, tunneling proceeds through a single level at $\epsilon_0=0$, occupied at zero temperature if $eV_g>0$ and empty otherwise, and the transmission element reduces to
    \begin{align}
        T_{\rm a, el}&(\varepsilon,V_g) =  \gamma_{\rm ua}^{\phantom{*}} \gamma_{\rm ad} \rho_F \e^{i 2\varepsilon a /\hbar v}  \frac{{-\Theta(V_{g})\e^{i\pi} + \Theta(-V_{g})}}{\varepsilon+eV_{g} + i 0^+} \ .  \label{Eq:Tael}
    \end{align}
We have arranged this expression so that its phase contributions are explicit: (i)~an exchange phase $\e^{i\pi}$, incurred when an operator of edge $\rm u$ is interchanged with one of edge $\rm d$ (see End Matter), equal to $-1$ for the occupied level and $+1$ for the empty one; (ii)~the Breit-Wigner phase generated as $\varepsilon+eV_g$ changes sign; (iii)~a sign $\mathrm{sgn}(-V_g)$, since the virtual intermediate state has positive energy for an empty antidot and negative energy for an occupied one (see End Matter);  and (iv)~a dynamical phase $\e^{i2\varepsilon a/\hbar v}$, nearly constant across a single resonance but governing the evolution between consecutive levels $n_0\to n_0+1$ through the distances $a$ and $\disAD$. For electrons, factors (i) and (iii) multiply to unity, leaving the Breit-Wigner phase (ii), so the $0\to\pi$ evolution carries no trace of statistics. For anyons, factor (i) becomes $\e^{i\pi\nu}$ and can no longer cancel (iii); instead the Breit-Wigner phase (ii) and the sign (iii) compensate, and the fractional exchange phase survives. The full calculation below confirms this and yields the power-law tunneling-in and tunneling-out densities of states of the fractional edge, integrated over the bias window.

\emph{Perturbative Keldysh calculation for anyons---}To compute the
interference current for Laughlin anyons, we bosonize the three channels and treat the antidot as a closed chiral edge of length $L$, with the anyonic statistics carried by Klein factors $\kappa_i$ (Green-function conventions and the contour defining the $\kappa_i$ are given in the End Matter). The contour choice~\cite{Altland.2015} fixes the Klein-factor exchange phases
\begin{align}
\begin{split}
    \kappa_{\rm u}\kappa_{\rm d}  = \kappa_{\rm d}\kappa_{\rm u} \e^{i\pi\nu}; \
    \kappa_{\rm a}\kappa_{\rm d}  = \kappa_{\rm d}\kappa_{\rm a} \e^{i\pi\nu}; \
    \kappa_{\rm u}\kappa_{\rm a} = \kappa_{\rm a}\kappa_{\rm u} \e^{i\pi\nu}
\end{split}
\end{align}
together with the rule that $\kappa_{i}\kappa_j=\kappa_j\kappa_i\e^{\pm i\pi\nu}$ implies $\kappa_{i}\bar{\kappa}_j=\bar{\kappa}_j\kappa_i\e^{\mp i\pi\nu}$. A perturbative expansion to third order in the tunneling couplings then yields the Aharonov-Bohm phase dependent part of the current in drain D2,
\begin{align}
        &I_{\rm d,\rm AB} = 2\frac{e^*|\gamma_{\rm ua}\gamma_{\rm ud}\gamma_{\rm ad}|}{\hbar^3} {\rm Im} \bigg\{ \e^{i\phi_{\rm AB}} \int_{-\infty}^\infty \td t_1  \int_{-\infty}^\infty \td t_2\,
        \notag \\ &\ \times
        \e^{i e^*\Delta V t_1/\hbar}\e^{i e^*(-V_{g}-\Delta V/2)t_2/\hbar}
        \notag\\ &\ \times
        \Big[ -\tilde{G}_{\rm d}^<(-a,-t_1) \tilde{G}^<_{\rm u}(a,t_2-t_1) \tilde{G}_{\rm a}^{\rm ret}(\disAD,-t_2)
        \notag \\ &\
        + \tilde{G}_{\rm d}^>(-a,-t_1) \tilde{G}_{\rm u}^>(a,t_2-t_1) \tilde{G}_{\rm a}^{\rm ret}(\disAD,-t_2) \Big] \bigg\} \ ,\label{Eq:Idretadv}
\end{align}
with $\Delta V$ the applied bias and $\phi_{\rm AB}$ the Aharonov-Bohm phase for a full loop. Equivalently $I_{\rm d, AB} = \frac{2e^*}{h}\,\frac{|\gamma_{\rm ua}\gamma_{\rm ud}\gamma_{\rm ad}|}{(e^*\Delta V)^2} \tilde{I}\cos(\phi_{\rm AB}+\delta  + \pi )$ with transmission phase $\delta(V_g)$ and amplitude $\tilde{I}(V_g)$, where the phase shift of $\pi$ amounts to the Breit-Wigner phase below resonance. For a short antidot the level spacing $\propto L^{-1}$ is large compared to other energy scales in the system and neglecting the oscillator modes, the antidot Green function reduces to
\begin{align}
    \tilde{G}_{\rm a}(x,t,\sigma_{12})   &\approx  \frac{1}{(2\pi)^\nu}   \e^{i\frac{2\pi\nu}{L}(N_a+(\sigma_{12}-1)/2)(x-vt)}
    \notag \\ &\ \ \times \e^{i\pi\nu{\rm sgn}(x) \frac{1-\sigma_{12}}{2}} \label{Eq:GFantidotWithNA} \ ,
\end{align}
the full expression being given in the End Matter~\cite{Geller.1997}.

The antidot anyons then behave as hard-core anyons~\cite{Averin.2007,Goldman.1995,Goldman.2005,Maasilta.1998,Kataoka.1999,DeLuca.2025}, with anyonic exchange statistics but fermionic exclusion.    
\begin{figure}
        \centering
        \includegraphics[width=1\linewidth]{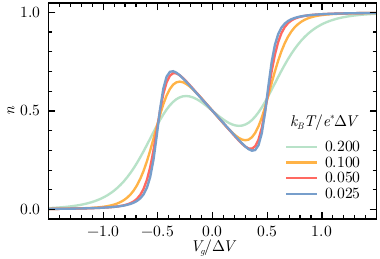}
        \caption{\label{fig:noneqoccupation} Non-equilibrium occupation $n$ of the antidot [Eq.~\eqref{Eq:nnoneq}] versus gate voltage $V_g$ at several temperatures, for balanced couplings $|\gamma_{\rm ua}|=|\gamma_{\rm ad}|=0.25\,e^*\Delta V$ (QPC transmission of about $10$--$30\%$). The level fills as $V_g$ increases, from empty at large negative $V_g$ to full at large positive $V_g$. At $V_g=-\Delta V/2$ it aligns with the u-edge chemical potential and couples strongly to that edge, so the occupation rises above $1/2$; at $V_g=0$ in- and out-tunneling balance, giving $n=1/2$; near $V_g=+\Delta V/2$ coupling to edge d dominates and $n$ dips below $1/2$ before saturating to unity. Temperature and
level broadening smooth the curve.}  
    \end{figure}

\emph{Phase evolution when tuning through a level---}We first focus on the case, $|e^*V_{g}|,\, e^*\Delta V \ll \Delta\varepsilon_{\rm a}$ with antidot level spacing $\Delta\varepsilon_{\rm a}$, for which tunneling is through a single antidot level, $\epsilon_{n_0}=0$, shifted by the plunger gate $V_{g}$.  Up to corrections $\mathcal{O}(a,L)$, we find
    \begin{align}
        &I_{\rm d,\rm AB} = \frac{2 e^*}{h} \frac{|\gamma_{\rm ua}\gamma_{\rm ud}\gamma_{\rm ad}|} {(8\pi^3\hbar^2\eta^{-2}v_{}^{2})^\nu}  \frac{4\pi^2 }{\GammaF^2(\nu)} \int_{-e^*\Delta V/2}^{e^*\Delta V/2}    \td \varepsilon
        \notag\\&\times
        {\rm Re} \left[\frac{ 
        {-\Theta(V_g) \e^{i\pi\nu} + \Theta(-V_g)}
        }{\varepsilon + e^*V_{g} +i0^+} \ \e^{i\phi_{\rm AB}}\right]
        \notag\\&\times
        |\varepsilon-e^*\Delta V/2|^{-1+\nu} |\varepsilon+e^*\Delta V/2|^{-1+\nu} \ . \label{Eq:CurrentAnyonsKeldyshSingleLevel}
    \end{align}  
    
Here, $\GammaF(x)$ is the gamma function and $\eta$ the short-distance ultraviolet cutoff. As anticipated from the $T$-matrix analysis, the phase difference between co-tunneling and direct tunneling is the exchange phase: the Breit-Wigner denominator $\varepsilon+e^*V_g$ persists by energy-time uncertainty, its phase and the relative sign canceling as for electrons [Eq.~\eqref{Eq:Tael}], while the tunneling densities of states fractionalize to $|\varepsilon - e^*V_i|^{-1+\nu}$~\cite{Kane.1992,Kane.1996,Kane.2003}. Using $(x+i0^+)^{-1}=\mathcal{P}x^{-1}-i\pi\delta(x)$, resonant tunneling adds a further phase of $\pi/2$ {due to the factor of $i$}. The exchange phase is read from the plateaus at $V_g<-\Delta V/2$ and $V_g>\Delta V/2$.

\emph{Non-equilibrium antidot occupation---}In the limit of small antidot extension $L$ with periodic boundary conditions, the anyons are effectively localized, causing Coulomb-blockade with occupation of the antidot $n(V_g)$  \cite{Averin.2007,Goldman.1995,Goldman.2005,Maasilta.1998,Kataoka.1999,DeLuca.2025}. Starting from Eq.~\eqref{Eq:Idretadv}, we perform a finite temperature calculation by using finite $T$ Green functions for upper and lower edges, and in the single level approximation ($\varepsilon_{n_0} = 0$), the antidot retarded Green function becomes
    \begin{align}
        \tilde{G}_{\rm a}^{\rm ret}(x,t) &= \frac{\Theta(t)}{(2\pi)^\nu} \left\{ \left[1-n(V_{g})\right] - \e^{i\pi\nu} n(V_{g})\right\}\ . \label{Eq:GretFiniteT}
    \end{align}
In a perturbative picture, where the occupation of the antidot is computed in the absence of the other edges, one finds a Fermi distribution $n(V_g) = n_F(-e^* V_g)$, in agreement with the zero temperature expression, Eq.~\eqref{Eq:CurrentAnyonsKeldyshSingleLevel}. We go beyond this result by taking into account the non-equilibrium occupation of the antidot due to coupling to the other edges \cite{Chamon.1993}. We obtain the gate voltage dependence of the antidot occupation from a higher order calculation of the current flowing from the edges $\alpha$ into the antidot \cite{Chamon.1993} 
    \begin{align}
        n(V_g) &= \int_{-\infty}^\infty \frac{ \sum_\alpha   |\gamma_{\alpha\rm a}|^2 G_{\alpha}^<\left(\varepsilon - e^*V_\alpha\right)  }{[\varepsilon+e^*V_g-\Lambda(\varepsilon)]^2 + [\Gamma(\varepsilon)/2]^2}\,\td \varepsilon\ , \label{Eq:nnoneq}
    \end{align}
where $V_\alpha = \pm \Delta V/2$ for $\alpha=\rm u$ and $\alpha=\rm d$, respectively. Here, the level broadening $\Gamma(\varepsilon)$ \cite{Kane.1996} and the level shift $\Lambda(\varepsilon)$ \cite{Jauho.1994} are obtained from a retarded self energy $\Sigma^{\rm ret}(\varepsilon) = \Gamma(\varepsilon)/2 + i\Lambda(\varepsilon)$, which can be defined for anyons similarly as for electrons by carefully treating the phase factors [see End Matter for details]. While according to a Fermi-distribution, the occupation changes only close to $V_g=0$ determined by the temperature, the  resulting non-equilibrium occupation [Fig.~\ref{fig:noneqoccupation}] changes quickly close to $V_g=\pm \Delta V/2$ due to the density of states anomaly of the edges at the chemical potential. 

For reasons of consistency, we then also incorporate the level broadening and level shift into the complex transmission amplitude used to obtain the Aharonov-Bohm phase dependent part of the current through the interferometer as
    \begin{widetext}
    \begin{align}
         I_{\rm d,\rm AB}&(V_g,T) =     \frac{2e^* |\gamma_{\rm ua}\gamma_{\rm ud}\gamma_{\rm ad}|}{\hbar^3} \ {\rm Re}  \Bigg\{ \e^{i\phi_{\rm AB}} \int_{-\infty}^\infty \td\varepsilon \frac{ \e^{i\frac{2a}{ \hbar v} \varepsilon   }}{(2\pi)^{\nu+1}}    
    \frac{[1-n(V_{g})]-\e^{i\pi\nu} n(V_{g})}{\varepsilon+e^*V_{g} - \Lambda(\varepsilon) +i\Gamma(\varepsilon)/2}
    \notag\\& \times
    \Bigg[   
    \e^{-\frac{e^*\Delta V}{\hbar v}\eta}
        G_{\rm d}^>\left(\varepsilon+\frac{e^*\Delta V}{2}\right) 
        G_{\rm u}^<\left(\varepsilon - \frac{e^*\Delta V}{2}\right)
    -  \e^{\frac{e^*\Delta V}{\hbar v}\eta}
        G_{\rm d}^<\left(\varepsilon + \frac{e^*\Delta V}{2}\right) 
        G_{\rm u}^>\left(\varepsilon-\frac{e^*\Delta V}{2}\right)
    \Bigg] \Bigg\}        \ . \label{Eq:FullPhaseTNonEq}
    \end{align} 
\end{widetext}

The computed phase evolution (upper panel of
Fig.~\ref{fig:3temperaturedependencelargeT}) interpolates between plateaus at $0$ and $\theta=\pi\nu$. This difference combines the denominator (Breit-Wigner) phase, which varies from $0$ to $\pi$, with the numerator phase, which runs from $0$ to $\pi\nu-\pi$ as direct tunneling gives way to co-tunneling.

Near the lead chemical potentials $V_g=\pm\Delta V/2$ the evolution is
non-monotonic, a feature that survives up to $k_BT\approx 0.2\,e^*\Delta V$. It arises from the interplay of three effects: resonant tunneling, enabled by the level broadening, which contributes a phase $\pi/2$; the rapid variation of the antidot occupation, which modulates the co-tunneling contribution with phase $\pi(\nu-1)$ in the numerator of Eq.~\eqref{Eq:FullPhaseTNonEq}; and the dip of the occupation below $1/2$ near $V_g=\Delta V/2$, which makes the phase overshoot  $\pi\nu$ before approaching the plateau value from below.

The exchange phase $\pi\nu$ is therefore read off slightly away from
$V_g=\pm\Delta V/2$, provided the level spacing is large compared to the bias. This requires the antidot charge to be screened by an external gate rather than by the quantum Hall edges; otherwise, by the Friedel sum rule, the electrostatic phase cancels the statistical one and the effect is unobservable (see Supplemental Material \cite{Supplement}). For $\Delta V=\SI{100}{\micro\volt}$, temperatures of $\SI{20}{\milli\kelvin}$ and $\SI{80}{\milli\kelvin}$ correspond to $k_BT/e^*\Delta V\approx 0.05$ and $0.2$, the largest value shown.

    \begin{figure}[t!]
        \centering
        \includegraphics[width=1\linewidth]{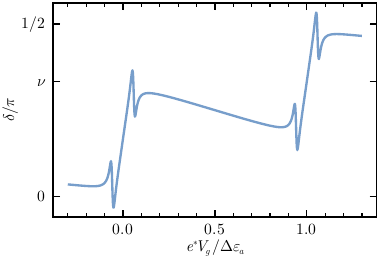}
        \caption{\label{fig:5tworesonances}Transmission phase $\delta$ versus gate voltage $V_g$ for tuning across two successive antidot levels, at $k_BT/e^*\Delta V=0.025$. The level spacing is $\Delta\varepsilon_{\rm a}$, so $e^*V_g/\Delta\varepsilon_{\rm a}=0$ and $1$ mark the two level crossings. Each crossing reproduces the non-monotonic behavior of Fig.~\ref{fig:3temperaturedependencelargeT}. Between crossings the phase depends on the antidot contact separation $\disAD$, here decreasing approximately linearly for $\disAD=L/4$.}  
    \end{figure} 

\textit{Validity}---The bare third-order result, Eq.~(\ref{Eq:CurrentAnyonsKeldyshSingleLevel}), diverges at $V_g=\pm\Delta V/2$, where the Breit--Wigner pole meets the power-law singularity of the anyonic tunneling density of states, so that plain perturbation theory fails there for arbitrarily weak coupling. The self-energy resummation into $\Gamma(\varepsilon)$ and $\Lambda(\varepsilon)$, together with the non-equilibrium occupation, Eq.~(\ref{Eq:nnoneq}), cures this divergence. As a consistency check, the modulus squared of the resummed transmission amplitude reproduces the sequential-tunneling result of Chamon and Wen  \cite{Supplement,Chamon.1993}. Contributions beyond this scheme are small as long as $r \equiv \Gamma(-e^*V_g)/\max\!\big(k_BT,\,\min_\alpha|e^*(V_g+V_\alpha)|\big) \ll 1$, the second argument being the detuning of the level from the nearest edge chemical potential. Parametrically, $r$ is the squared renormalized coupling at the infrared scale. This condition holds for all temperatures shown in the plateau regions, from which the exchange phase is extracted. Near $V_g=\pm\Delta V/2$ it requires $k_BT$ to exceed the strong-coupling scale of the antidot couplings; at lower temperatures the lineshape there is qualitative rather than
quantitative.

\emph{Tuning between two levels---}The phase evolution between two consecutive levels depends on the dynamical phases, and thus on the geometry, i.e., the distance between the QPCs on the edges, $a$, and in the antidot  $\disAD$ (for a discussion on corrections for finite $a$, see \cite{Supplement}). We show the phase evolution for tuning from a level at zero to the next level at $\Delta \varepsilon_{\rm a}$ in Fig.~\ref{fig:5tworesonances}, assuming the levels lie far enough apart that they contribute individually and independently to the interference current \cite{Supplement} { and that the dynamical phase between the levels differs by $\e^{i \Delta \varepsilon_{\rm a} b/\hbar v_{\rm a}}$}. The result shows that such a measurement is not suitable to determine the exchange phase.

\emph{Conclusion}---We have developed a systematic non-equilibrium theory of the interference current  for the modified Fabry-Perot interferometer proposed in Ref.~\cite{Kivelson.2024a}. A perturbative Keldysh calculation shows that the transmission-phase difference between co-tunneling through an occupied level and tunneling through an empty level is the exchange phase $\pi\nu$, carried by the Klein factors of the anyon fields. Including the non-equilibrium occupation of the antidot and the level broadening from the anyonic density-of-states anomaly, tuning through a resonance produces a non-monotonic evolution  of the transmission phase, in contrast to the monotonic $0\to\pi$ evolution for electrons. This evolution appears in the Aharonov-Bohm oscillations of the current and is robust against finite temperature and small changes in the device geometry. The exchange phase is encoded in the difference between the resulting plateaus and is best extracted from a single resonance, since tuning between consecutive levels is dominated by geometry-dependent dynamical phases. Finally, the statistical phase is observable only when the antidot charge is screened by an external gate; otherwise the Coulomb phase from the displaced charge cancels it.  Extending this analysis to hierarchical and non-Abelian states is a natural next step.

\acknowledgements 
We acknowledge  helpful conversations with B.I. Halperin, J. Ehrets, and P. Kim. This research was  supported by Deutsche Forschungsgemeinschaft (DFG) under the Grant No.\@ 406116891 within the Research Training Group RTG 2522/1.

%

\onecolumngrid \vspace{0.5cm}
\begin{center}{\large\textbf{End Matter}}\end{center} \vspace{0.1cm}
\twocolumngrid \noindent
\emph{Details on $T$-matrix calculation---}In the case of electrons, the calculation simplifies as we can introduce  antidot single-particle energy levels $\epsilon_m$ and operators $d_m^\dagger$, which create an electron in the dot eigenstate $m$. 
     Thus, the dot Hamiltonian becomes  $H_{a} = \sum_m (\epsilon_m + e V_{g}) d_m^\dagger d_m$.  
	 We assume that tunneling is predominately through a single level $\epsilon_0=0$.  
     We obtain the complex transmission amplitude at energy $\varepsilon$, between an initial state $\ket{i} $ and a final state $\ket{f} $ through the antidot level with occupation $n_{\rm a}$ via a virtual  intermediate state. 
     Operators $c_\alpha^\dagger(\varepsilon)$ ($c_\alpha(\varepsilon)$) create (annihilate) an electron at energy $\varepsilon$ in edge $\alpha$,  $\psi_\alpha^\dagger(x)$ creates an electron at position $x$ in lead $\alpha$, and $d$ annihilates an electron in the antidot. 
	The complex transmission matrix element for such a process can be obtained from the $T$-matrix formalism as 
	\begin{align}
        T^{n_{\rm a}}_{\rm a, el}(\varepsilon) &= \sum_{m} \bra{f} H_{\rm tun} \ket{m}  \frac{1}{\varepsilon- E_m + i 0^+} \bra{m}H_{\rm tun}\ket{i}\, \rho_F\ , \label{Eq:TaTmatrixElectrons} 
	\end{align}
	where $\rho_F$ is the density of states at the Fermi level in edges $\rm u$ and $\rm d$, $n_{\rm a}$ is the occupation of the antidot level, and $+i0^+$ is the regularization for the retarded Green function, which in the presence of many levels to higher order may be replaced by a level broadening term. 
	For direct tunneling through an empty level, the initial state is $\ket{i}=c^\dagger_{\rm u}(\varepsilon) \ket{0}$ with energy $E_i=\varepsilon$ and the final state is $\ket{f}=c^\dagger_{\rm d}(\varepsilon) \ket{0}$. The electron jumps from the upper edge into the antidot forming an intermediate state with energy $E_m=-e^*V_g$. Thus
	\begin{align}
        T^{n_{\rm a }=0}_{\rm a, el}(\varepsilon) &= 
        \bra{0} c_{\rm d}(\varepsilon)
        (-\gamma_{\rm ad})\psi_{\rm d}^\dagger(-a) d
        \frac{1}{\varepsilon + eV_g + i0^+}
        \notag \\ & \ \ \ \ \ \ \times
        (-\gamma_{\rm ua})d^\dagger \psi_{\rm u}(a)
        c_{\rm  u}^\dagger(\varepsilon)
        \ket{0} \nonumber \\ 
        &= \frac{\gamma_{\rm ad}\gamma_{\rm ua} \e^{2i\varepsilon a}}{\varepsilon + eV_g + i0^+} \ .
	\end{align} 
    Here, the dynamical phase arises when inserting the Fourier representation for the field operators.

	For co-tunneling through an occupied level, initial and final states are $\ket{i}=d^\dagger  c^\dagger_{\rm u}(\varepsilon) \ket{0}$ and $\ket{f}=  d^\dagger c^\dagger_{\rm d}(\varepsilon) \ket{0}$. Here, an electron first jumps out of the antidot to the lower edge to form an intermediate state. Again, only a single combination of tunneling terms yields a non-vanishing expectation value, where the denominator becomes $(\varepsilon -eV_g -i0^+ ) - (2\varepsilon)$ such that
	\begin{align}
        &T^{n_{\rm a }=1}_{\rm a, el}(\varepsilon) = 
        \bra{0} c_{\rm d}(\varepsilon) \, d\,
        (-\gamma_{\rm ua}) d^\dagger \psi_{\rm u}(a) 
        \frac{1}{-\varepsilon - eV_g - i0^+}
        \notag \\ & \ \ \ \ \ \ \times
        (-\gamma_{\rm ad})\psi_{\rm d}^\dagger(-a) \,d\,
        d^\dagger c_{\rm  u}^\dagger(\varepsilon) 
        \ket{0}
\nonumber        \\
        &= \frac{\bra{0}c_{\rm d}(\varepsilon) \psi_{\rm u}(a) \psi_{\rm d}^\dagger(-a) c_{\rm  u}^\dagger(\varepsilon) \ket{0}\gamma_{\rm ua}\gamma_{\rm ad}  }{-\varepsilon - eV_g - i0^+} \nonumber \\ 
        &= \frac{\e^{i\pi}\gamma_{\rm ua}\gamma_{\rm ad} \e^{2i\varepsilon a}}{-\varepsilon - eV_g - i0^+} 
         = \frac{-\e^{i\pi}\gamma_{\rm ua}\gamma_{\rm ad} \e^{2i\varepsilon a}}{\varepsilon + eV_g + i0^+} \ . 
	\end{align}
	
	Here the $\e^{i\pi}=-1$ factor is due to an exchange of $\psi_{\rm u}(a)$ and $\psi_{\rm d}^\dagger(-a)$ operators.
    Within perturbation theory,   the level is occupied if it lies below zero, $-eV_g<0$, and empty otherwise. We 
	thus find the complex transmission amplitude as given in Eq.~\eqref{Eq:Tael}.

\emph{Choice of Klein factors---} The Klein factors can be defined by connecting the edges to a single closed contour such that no tunneling points cross and the chirality is respected \cite{Guyon.2002,Altland.2015}. Choosing an origin on this contour then uniquely defines the Klein factors by the boson field $\phi(x)$ commutation relations on the connected edge, however, observables do not depend on this choice. As the exchange phases are a topological property of anyons in the fractional quantum Hall state, the Klein factors are not  affected if we adiabatically deform the antidot edge contour, as long as we do not introduce a crossing of tunneling points. We can therefore elongate the edge toward the outside of the interferometer (to the right in Fig.~\ref{fig:1setup}) to infinity. This gives rise to three infinite edges for which a unique closed contour is known \cite{Altland.2015} by connecting edge u to the antidot at 0 and after going through the antidot edge, connecting it to edge d, which then is connected to edge u again. For the topology it does not matter if the antidot edge is closed or open at infinity. As described in the main text,  we  put  the zero point on the start of contour u and the direction according to the chirality. This gives rise to Klein factor exchange phases  $
    \kappa_i \kappa_j = \kappa_j \kappa_i\, e^{i\pi\nu\alpha_{ij}}$ and $ \overline{\kappa}_i \kappa_j= \kappa_j  \overline{\kappa}_i$ \,$ e^{-i\pi\nu\alpha_{ij}}$ 
with 
\begin{align}
    \alpha= \left( \begin{matrix}
        0 & 1 & 1  \\
        -1 & 0 & 1   \\
        -1 & -1 & 0     
    \end{matrix}\right) \ .
\end{align}
Here $i=1$ corresponds to the  upper edge $\rm u$, $i=2$ to the antidot $\rm a$, and $i=3$ to the lower edge $\rm d$.

\emph{Details on Keldysh calculation---} 
We bosonize the three channels and introduce field operators
$\psi_{\rm a}(x,t)=\kappa_{\rm a}\e^{ie^*V_{g} t }\e^{i\phi_{\rm a}(x,t)} / (2\pi)^{\nu/2}$ on the antidot  with
\begin{align}
\phi_{\rm a}(x,t) &=  -i\chi_0 + i \frac{\pi\nu}{L}(2N-1 )(x-vt) + i\phi_{\rm osc}(x-vt) \ ,
\end{align}
where $[\chi_0,N]=i$, the displacement field obeys
$[\phi_{j}(x),\phi_{j}(y)]=i\pi\nu\,\text{sgn}(x-y)$, and $\phi_{\rm osc}$ are
the oscillator plasmon modes. The Klein factor dressed Green functions are
\begin{align}
    \tilde{G}_i(x,t,\sigma_{12}) &= \e^{i\pi\nu{\rm sgn}(x)\frac{1-\sigma_{12}}{2}} G_i(x,t,\sigma_{12})\\
    G_i(x,t,\sigma_{12}) &= \ex{\e^{i\phi_i(\sigma_{12}x,\sigma_{12}t)}\e^{-i\phi_i(0,0)}} \ ,
\end{align}
with expectation values under the free Hamiltonian at $V_g=0$, Keldysh indices
$\sigma_{12}=[\sigma_2-\sigma_1 + {\rm sgn}(t)(\sigma_2+\sigma_1)]/2$ (so
$\sigma_{12}=\pm$ give the greater and lesser functions) and
$\tilde{G}^{\rm ret}(x,t) = \theta(t) [\tilde{G}^>(x,t) - \tilde{G}^<(x,t)]$.
The prefactor encodes the Klein-factor exchange phases~\cite{Guyon.2002,Altland.2015}.
Using the boson fields, the full zero-temperature antidot Green function is~\cite{Geller.1997}
\begin{align}
    &\tilde{G}_{\rm a}(x,t,\sigma_{12})
     = \frac{1}{(2\pi)^\nu}   \e^{i\frac{\pi\nu}{L}(2N_a-1+\sigma_{12} )(x-v_{\rm a}t)}
    \notag \\ &\times
    \e^{i\pi\nu{\rm sgn}(x) \frac{1-\sigma_{12}}{2}} \left[ \frac{\sinh\left(\frac{\pi\eta}{L}\right) \e^{-i\frac{\pi}{L}\sigma_{12}(x-v_{\rm a}t)}}{\sinh\left( \frac{ \pi}{L}(\eta-i\sigma_{12}(x-v_{\rm a}t)\right)}\right]^{\nu} \ , \label{Eq:GantidotFull}
\end{align}
with $x=x_2-x_1$ where $x_1,x_2\in[0,L]$. Neglecting the oscillator modes yields Eq.~\eqref{Eq:GFantidotWithNA} of the main text.

We compute the current in drain D2 via tunneling currents at the QPCs. Here the tunneling current operator between edges $n$ and $m$ is given by  
    \begin{align}
        I_{nm}(t) &= \frac{ie^*}{\hbar} \left[O^+_{nm}(t)
        + \rm h.c.\right] \\ 
        O^+_{nm}(t) &= \gamma_{nm}\psi_m^\dagger(x_{mn},t)\psi_n(x_{nm},t)\ .
    \end{align}
At zero temperature and for tunneling through a single level, we obtained  in the main text Eq.~\eqref{Eq:Idretadv} for the current to third order in the tunneling couplings.
For the Green functions on the edges at 
finite temperature $T$, we use 
\begin{align}
    \tilde{G}_{\rm u/d}(x,t,&\tilde{\sigma}_{12}) = \e^{i\pi\nu{\rm sgn}(x)\frac{1-\tilde{\sigma}_{12}}{2}}
    \notag\\ \times&
        \left[\frac{k_BT \eta / v\hbar}{2\sin\left[\pi k_BT \left(\eta-i\tilde{\sigma}_{12} (x-vt)\right)/v\hbar\right]}\right]^\nu 
\end{align}
with $\tilde{\sigma}_{12}    = ({\sigma_2-\sigma_1 + {\rm sgn}(x)(\sigma_2+\sigma_1)})/{2}$. We find the Fourier representation of the bosonic greater Green function to be  
\begin{align}
   &G_{\rm u/d}^>(\omega) = 2\hbar\GammaF(1-\nu) \left(\frac{2\pi k_BT \eta}{\hbar v}\right)^\nu  
   \notag \\ &\times 
   \mathrm{Re}\left[ \frac{(i)^{-\nu}\GammaF\!\left(\frac{\nu}{2}- i \frac{\hbar \omega}{2\pi k_B T} + 1 \right)}{\GammaF\!\left(-\frac{\nu}{2}-i\frac{\hbar \omega}{2\pi k_B T} + 1\right) \left(\pi\nu k_B T - i \hbar \omega\right)} \right] \ , \intertext{}\notag
\end{align}
where $ \GammaF(y)$ is the gamma function (not to be confused with the level broadening $\Gamma(\varepsilon)$), and $G^<(\omega)=G^>(-\omega)$. 
Using these Green functions together with Eq.~\eqref{Eq:GFantidotWithNA} in the expression for the current, Eq.~\eqref{Eq:Idretadv}, we find the following zero temperature expression for $V_g>0$ when tuning from level 0 to the first empty level for $\Delta V>0$ 
    \begin{widetext}
    \begin{align}
     &I_{\rm d,\rm AB}(V_g,T=0) = \frac{2 e^* }{h} \frac{|\gamma_{\rm ua}\gamma_{\rm ud}\gamma_{\rm ad}|} {(8\pi^3\hbar^2\eta_{}^{-2}v^2)^\nu}  \frac{4\pi^2}{\GammaF^2(\nu)} {\rm Re}\Bigg\{ \e^{i\phi_{\rm AB}} \int_{-e^*\Delta V/2}^{e^*\Delta V/2}    {\td\varepsilon}  |\varepsilon - e^*\Delta V/2|^{-1+\nu} |\varepsilon+e^*\Delta V/2|^{-1+\nu} \e^{i2a\varepsilon/\hbar v} 
        \notag\\ &\ \times {\rm sgn}(-V_{g})
        \left[\frac{\e^{i\pi\nu [{1-{\rm sgn}(-V_{g})}]/{2}}}{\varepsilon + e^*V_{g} +  i0^+}-\frac{\e^{-i\frac{2\pi\nu}{L}{\rm sgn}(-V_{g}) \disAD} \e^{i\pi\nu [{1+{\rm sgn}(-V_{g})}]/{2}}}{\varepsilon + e^*V_{g} + {\rm sgn}(-V_{g}) \frac{2\pi\nu}{L}\hbar v_{\rm a}  +  i 0^+}\right] \Bigg\} + \mathcal{O}(L,\gamma^3)\ . 
\end{align}
\end{widetext}

\noindent For the numerical evaluations, we measure length  in units of $\hbar v/e^*\Delta V$ and energies in units of $e^*\Delta V$. We choose $\eta=10^{-4}$  for all numerical calculations and for the case  where more than one level is considered in Fig.~\ref{fig:5tworesonances}, we set the length of the antidot to $ L =0.25\hbar v/e^*\Delta V$. For parameters $v\approx \SI{e5}{\meter\per\second}$ and $\Delta V\approx \SI{100}{\micro\volt}$, the characteristic length scale takes the value $\hbar v/e^*\Delta V\approx\SI{2}{\micro\meter}$.

\emph{Details on non-equilibrium occupation---}We obtain the self-energy due to integrating out the lead edges as $\Sigma_\alpha^{\rm ret}(\varepsilon) = |\gamma_{\alpha a}|^2G^{\rm ret}_\alpha(\varepsilon)$ \cite{Jauho.1994}, which determines the level broadening and level shift as the real and imaginary part. For anyons, we use a convention for the Green functions that differ from the typical electron definitions by the phase factors $\pm i$. Starting from the real greater and lesser Green function, we can define the transformation to the retarded one as 
\begin{align}   
    G^{\rm ret}_\alpha(\varepsilon) &= \frac{1}{2} \left[G_\alpha^>(\varepsilon) + G_\alpha^<(\varepsilon)\right] 
    \notag\\
    &\ + \frac{i}{2} \cot\left(\frac{\pi\nu}{2}\right) \left[G_\alpha^>(\varepsilon) - G_\alpha^<(\varepsilon)\right] \ . 
\end{align}
This is a consistent definition in the sense that the lesser Green function is exactly a product of the spectral function $A_\alpha(\varepsilon) = 2\mathrm{Re}(G^{\rm ret}_\alpha(\varepsilon))$ and a Fermi-Dirac distribution, i.e., $G^<(\varepsilon) = A(\varepsilon) n_F(\varepsilon)$ and $G^>(\varepsilon) = A(\varepsilon) [1-n_F(\varepsilon) ]$. The occupation of the dot is then determined from an integral over a non-perturbative lesser Green function of the dot in the presence of the lead edges \cite{Chamon.1993}, which by the Dyson equation is the integrand of Eq.~\eqref{Eq:nnoneq}. {The level broadening is thus $\Gamma(\epsilon)=2\sum_\alpha |\gamma_{\alpha \rm a}|^2 \mathrm{Re} G_\alpha^{\rm ret}(\omega + e^*V_\alpha)$, and the level shift is $\Lambda(\epsilon)=\sum_\alpha |\gamma_{\alpha \rm a}|^2 \mathrm{Im} G_\alpha^{\rm ret}(\omega + e^*V_\alpha)$.} These definitions are consistent with the results by  \citet{Chamon.1993} for resonant anyon tunneling and extend them by including virtual processes (see Supplemental Material \cite{Supplement}).
\ifarXiv
\clearpage 
\onecolumngrid
    \foreach \x in {1,...,\numbersupplementpages}
    {   
        \includepdf[pages={\x}]{\supplementfilename}
    }
\fi 
\end{document}